\documentclass [12 pt]{article}
\def\be{\begin{equation}}
\def\eea{\end{eqnarray}}
\def\bea{\begin{eqnarray}}
\def\ee{\end{equation}}

\usepackage[psamsfonts]{amssymb}
\usepackage{amsmath}
\author{F. Kheirandish$^{1}$ \footnote{fardin$_{-}$kh@phys.ui.ac.ir} and M.
Amooshahi$^{1}$ \footnote{amooshahi@sci.ui.ac.ir}
\\ $^{1}$ {\small Department of Physics, University of Isfahan,}
\\ {\small Hezar Jarib Ave., Isfahan, Iran.}}
\title{A minimal coupling method for dissipative quantum systems}
\begin{document}
\maketitle
\begin{abstract}
\noindent Quantum dynamics of a general dissipative system
investigated by its coupling to a Klein-Gordon type field as the
environment by introducing a minimal coupling method. As an
example, the quantum dynamics of a damped three dimensional
harmonic oscillator investigated and some transition
probabilities indicating the way energy flows between the
subsystems obtained. The quantum dynamics of a dissipative two
level system considered.
\end{abstract}
\section{Introduction}
In classical mechanics dissipation can be taken into account by
introducing a velocity dependent damping term into the equation
of motion. Such an approach is no longer possible in quantum
mechanics where a time-independent Hamiltonian implies energy
conservation and accordingly we can not find a unitary time
evolution operator for both states and observable quantities consistently.\\
To investigate the quantum mechanical description of dissipating
systems, there
 are some treatments, one can consider the interaction
between two systems via an irreversible energy flow [1,2], or take
a phenomenological treatment for a time dependent Hamiltonian
which describes damped oscillations, here we can refer the
interested reader to Caldirola-Kanai Hamiltonian for a damped
harmonic oscillator [3].
\begin{equation}\label{dm1}
H(t)=e^{-2\beta t}\frac{p^2}{2m}+e^{2\beta t}\frac{1}{2}m\omega^2
q^2.
\end{equation}
There are significant difficulties about the quantum mechanical
solutions of the Caldirola-Kanai Hamiltonian, for example
quantizing such a Hamiltonian violates the uncertainty relations
or canonical commutation rules and the uncertainty relations
vanish
as time tends to infinity.[4,5,6,7,8] \\
In 1931, Bateman [9] presented the mirror-image Hamiltonian which
consists of two different oscillator, where one of them
represents the main one-dimensional damped harmonic oscillator.
Energy dissipated by the main oscillator completely will be
absorbed by the other oscillator and thus the energy of the total
system is conserved. Bateman Hamiltonian is given by
\begin{equation}\label{dm2}
H=\frac{p\bar{p}}{m}+\frac{\beta}{2m}(\bar{x}\bar{p}-xp)+(k-\frac{\beta^2}{4m})x\bar{x},
\end{equation}
with the corresponding Lagrangian
\begin{equation}\label{dm3}
L=m\dot{x}\dot{\bar{x}}+\frac{\beta}{2}(x\dot{\bar{x}}-\dot{x}\bar{x})-kx\bar{x},
\end{equation}
canonical momenta for this dual system can be obtained from this
Lagrangian as
\begin{equation}\label{dm3}
p=\frac{\partial L}{\partial
\dot{x}}=m\dot{\bar{x}}-\frac{\beta}{2}\bar{x},\hspace{1.50
cm\bar{p}}=\frac{\partial L}{\partial
\dot{\bar{x}}}=m\dot{x}+\frac{\beta}{2}x,
\end{equation}
dynamical variables $ x, p $ and $\bar{p},\bar{x} $ shoud satisfy
the commutation relations
\begin{equation}\label{dm4}
[x,p]=i,\hspace{2.00 cm} [\bar{x},\bar{p}]=i,
\end{equation}
however the time-dependent uncertainty products obtained in this
way, vanishes as time tends to infinity.[10]\\
Caldirola [3,11] developed a generalized quantum theory of a
linear dissipative system in 1941 : equation of motion of a single
particle subjected to a generalized non conservative force $ Q $
can be written as
\begin{equation}\label{dm5}
\frac{d}{d t}(\frac{\partial T}{\partial \dot{q}})-\frac{\partial
T }{\partial q}=-\frac{\partial V}{\partial q}+Q(q),
\end{equation}
where $ Q_r =-\beta(t)\sum a_{r j}\dot{q}_j $, and $ a_{rj} $'s
are some constants, changing the variable $t$ to $t^{*}$, using
the following nonlinear transformation
\begin{equation}\label{dm6}
t^*=\chi(t),\hspace{1.50cm}dt=\phi(t)dt^*,\hspace{1.50cm}\phi(t)=e^{\int_0^t\beta(t')dt'},
\end{equation}
 together with the definitions
\begin{equation}\label{dm7}
  \dot{q}^*=\frac{d q}{d t^*},\hspace{1.00cm}
L^*=L(q,\dot{q}^*,t^*),\hspace{1.00cm}p^*=\frac{\partial
L^*}{\partial \dot{q}^*},
\end{equation}
the Lagragian equations, can be obtained from
\begin{equation}\label{dm8}
\frac{d}{d t^*}(\frac{\partial L^*}{\partial
\dot{q}^*})-\frac{\partial L^*}{\partial q}=0.
\end{equation}
where $ H^*=\sum p^*\dot{q}^*-L^* $. Canonical commutation rule
and Schrodinger equation in this formalism are
\begin{equation}\label{dm9}
[q,p^*]=i,\hspace{2.00cm}H^*\psi=i\frac{\partial \psi}{\partial
t^*},
\end{equation}
but unfortunately uncertainty relations vanish as time goes to
infinity.[10]\\
Perhaps one of the effective approaches in quantum mechanics of
dissipative systems is the idea of considering an environment
coupled to the main system and doing calculations for the total
system but at last for obtaining observables related to the main
system, the environment degrees of freedom must be eliminated.
The interested reader is referred to the Caldeira-Legget model
[12,13]. In this model the dissipative system is coupled with an
environment made by a collection of $ N $ harmonic oscillators
with masses $ m_n $ and frequencies $\omega_n $, the interaction
term in Hamiltonian is as follows
\begin{equation}\label{dm10}
H'=-q\sum_{n=1}^N c_n
x_n+q^2\sum_{n=1}^N\frac{c_n^2}{2m_n\omega_n^2},
\end{equation}
where  $ q $ and $ x_n $ denote coordinates of system and
environment respectively and the constants $ c_n $ are
called coupling constants.\\
The above coupling is not suitable for dissipative systems
containing a dissipative term proportional to velocity. In fact
with above coupling we can not obtain a Heisenberg equation like $
\ddot{q}+\omega^2 q+\beta\dot{q}=\xi( t)$, for a damped harmonic
oscillator, consistently. In this paper we generalize the
Caldeira-Legget model to an environment with continuous degrees
of freedom by a coupling similar to the coupling between a charged
particle and the  electromagnetic field known as the minimal
coupling. In section 2, the idea of a generalized minimal coupling
method is introduced and in section 3 the quantum dynamics of a
three dimensional damped harmonic oscillator is investigated. In
section 4, we study quantum dynamics of a dissipative  two level
system. Finally in section 5, dynamics of quantum field of
reservoir is investigated.
\section{Quantum dynamics of a dissipative system}
Quantum mechanics of a dissipative system  can be investigated by
introducing a reservoir or an environment that interacts with the
system through a new kind of minimal coupling term. For this
purpose let the damped system be a particle with mass $ m $ under
an external potential $ v(\vec{x}) $. The total Hamiltonian,
i.e., system plus environment, is
\begin{equation} \label{d1}
H=\frac{(\vec{p}-\vec{R})^2}{2m}+v(\vec{x})+H_B,
\end{equation}
where  $ \vec{x} $ and $ \vec{p} $ are position and canonical
conjugate momentum operators of the particle respectively and
satisfy the canonical commutation rule
\begin{equation}\label{d2}
[\vec{x},\vec{p}]=i,
\end{equation}
 $ H_B $ is the reservoir Hamiltonian defined by
\begin{equation}\label{d3}
H_B(t)=\int_{-\infty}^{+\infty}d^3k  \omega_{\vec{k}}
b_{\vec{k}}^\dag(t) b_{\vec{k}}(t), \hspace{1.50 cm}
\omega_{\vec{k}}=|\vec{k}|.
\end{equation}
Annihilation and creation operators $ b_{\vec{k}}$,
$b_{\vec{k}}^\dag $, in any instant of time, satisfy the following
commutation relations
\begin{equation}\label{d4}
[b_{\vec{k}}(t),b_{\vec{k}'}^\dag(t)]=\delta(\vec{k}-\vec{k}'),
\end{equation}
 and we will show later that reservoir is
a Klein-Gordon type equation with a source term. Operator $
\vec{R} $ have the basic role in interaction between the system
and reservoir and is defined by
\begin{equation}\label{d5}
\vec{R}(t)=\int_{-\infty}^{+\infty}d^3k [f(\omega_{\vec{k}})
b_{\vec{k}}(t)+f^*(\omega_{\vec{k}})b_{\vec{k}}^\dag(t)]\vec{k},
\end{equation}
let us call the function $ f(\omega_{\vec{k}}) $, the coupling
function. It can be shown easily that Heisenberg equation for $
\vec{x}(t) $ and $\vec{p}$ leads to
\begin{eqnarray}\label{d6}
&&\dot{\vec{x}}=i[H,\vec{x}]=\frac{\vec{p}-\vec{R}}{m},\nonumber\\
&& \dot{\vec{p}}=i[H,\vec{p}]=-\vec{\nabla}v,
\end{eqnarray}
where after omitting $\vec{p}$, gives the following equation  for
the damped system
\begin{equation}\label{d6.5}
m\ddot{\vec{x}}=-\vec{\nabla}v-\dot{\vec{R}}.
\end{equation}
 Using (\ref{d4}) the Heisenberg equation for $ b_k$, is
\begin{equation}\label{d7}
\dot{b}_{\vec{k}}=i[H,b_{\vec{k}}]=-i\omega_{\vec{k}}
b_{\vec{k}}+if^*(\omega_{\vec{k}})\vec{k}.\dot{\vec{x}},
\end{equation}
with the following formal solution
\begin{equation}\label{d8}
b_{\vec{k}}(t)=b_{\vec{k}}(0)e^{-i\omega_{\vec{k}}
t}+if^*(\omega_{\vec{k}})\vec{k}. \int_0^t d t'
e^{-i\omega_{\vec{k}}(t-t')} \dot{\vec{x}}(t'),
\end{equation}
substituting $ b_{\vec{k}}(t) $ from (\ref{d8}) into (\ref{d6.5}),
one obtains
\begin{eqnarray}\label{d9}
&&m\ddot{\vec{x}}+\int_0^t d
t'\dot{\vec{x}}(t')\gamma(t-t')=-\vec{\nabla}v+\vec{\xi}(t)\nonumber\\
&&\gamma(t)=\frac{8\pi}{3}\int_0^\infty d\omega_{\vec{k}}
|f(\omega_{\vec{k}})|^2\omega_{\vec{k}}^5\cos\omega_{\vec{k}} t\nonumber\\
&&\vec{\xi}(t)=i\int_{-\infty}^{+\infty} d^3 k
\omega_{\vec{k}}(f(\omega_{\vec{k}})b_{\vec{k}}(0)
e^{-i\omega_{\vec{k}}t}-f^*(\omega_{\vec{k}})b_{\vec{k}}^\dag(0)e^{i\omega_{\vec{k}}t})\vec{k}.
\end{eqnarray}
It is clear that the expectation value of $ \vec{\xi}(t) $ in any
eigenstate of $ H_B $, is zero. For the following special choice
of coupling function
\begin{equation}\label{d10}
f(\omega_{\vec{k}})=\sqrt{\frac{3\beta}{4\pi^2\omega_{\vec{k}}^5}},
\end{equation}
equation (\ref{d9}) takes the form
\begin{eqnarray}\label{d11}
&&m\ddot{\vec{x}}+\beta\dot{\vec{x}}=-\vec{\nabla}v+\vec{\tilde{\xi}}(t)\nonumber\\
&&\vec{\tilde{\xi}}(t)=i\sqrt{\frac{3\beta}{4\pi^2}}\int_{-\infty}^{+\infty}
\frac{d^3 k}{\sqrt{\omega_{\vec{k}}^3}} (
b_{\vec{k}}(0)e^{-i\omega_{\vec{k}}t}-
b_{\vec{k}}^\dag(0)e^{i\omega_{\vec{k}}t}).
\end{eqnarray}
In the following we investigate the quantum dynamics of a two
level system and an isotropic three dimensional harmonic
oscillator, both interacting with a reservoir, as prototypes of
dissipative models.
\section{Quantum dynamics of a three dimensional damped harmonic
oscillator}
\subsection{Quantum dynamics}
 For a three dimensional harmonic oscillator with mass $ m $ and frequency $ \omega $ we have $
 v(\vec{x})=\frac{1}{2}m^2\omega^2\vec{x}^2 $ and therefore we can
 write (\ref{d11}) as
 \begin{equation}\label{d11.01}
 \ddot{\vec{x}}+\frac{\beta}{m}\dot{\vec{x}}+\omega^2\vec{x}=\frac{\vec{\tilde{\xi}}(t)}{m},
\end{equation}
with the following solution
\begin{eqnarray}\label{d11.02}
&&\vec{x}(t)=e^{-\frac{\beta t}{2m}}(\vec{\hat{A}}e^{i\omega_1
t}+\vec{\hat{B}}
e^{-i\omega_1 t})+\vec{M}(t),\nonumber\\
&& \vec{M}(t)=i\int_{-\infty}^{+\infty}d^3 k
\sqrt{\frac{3\beta}{4\pi^2m^2\omega_{\vec{k
}}^3}}[\frac{b_{\vec{k}}(0)}{\omega^2-\omega_{\vec{k}}^2-i\frac{\beta}{m}
\omega_{\vec{k}}}e^{-i\omega_{\vec{k}}
t}-\frac{b_{\vec{k}}^\dag(0)}{\omega^2-\omega_{\vec{k}}^2+\frac{i\beta}{m}
\omega_{\vec{k}}}e^{i\omega_{\vec{k}}t}]\vec{k},\nonumber\\
&&
\end{eqnarray}
where $ \omega_1=\sqrt{\omega^2-\frac{\beta^2}{4m^2}}$. Operators
$ \vec{\hat{A}} $ and $ \vec{\hat{B}} $, are specified by initial
conditions
\begin{eqnarray}\label{d11.03}
\vec{\hat{A}}+\vec{\hat{B}}&=&\vec{x}(0)-\vec{M}(0),\nonumber\\
(\frac{-\beta}{2m}+i\omega_1)\vec{\hat{A}}+
(\frac{-\beta}{2m}-i\omega_1)\vec{\hat{B}}&=&
\dot{\vec{x}}(0)-\dot{\vec{M}}(0)\nonumber\\
&=&\frac{\vec{p}(0)-\vec{R}(0)}{m}-\dot{\vec{M}}(0),
\end{eqnarray}
solving above equations and substituting $ \vec{\hat{A}}$ and $
\vec{\hat{B}}$ in (\ref{d11.02}), one obtains
\begin{eqnarray}\label{d11.04}
&&\vec{x}(t)=e^{-\frac{\beta t}{2m}}
\{\frac{\vec{p}(0)}{m\omega_1}\sin\omega_1 t
+\vec{x}(0)\cos\omega_1 t+
\frac{\beta}{2m\omega_1}\vec{x}(0)\sin\omega_1 t\nonumber\\
&&-\frac{\vec{R}(0)}{m\omega_1}\sin\omega_1 t -\frac{\beta
}{2m\omega_1}\vec{M}(0)\sin\omega_1t-\vec{M}(0)\cos\omega_1 t
-\frac{1}{\omega_1}\dot{\vec{M}}(0)\sin\omega_1 t\}+\vec{M}(t),\nonumber\\
&&
\end{eqnarray}
also substituting $ \vec{x}(t) $ from (\ref{d11.04}) in (\ref{d8})
we can find a stable solution for $ b_{\vec{k}}(t) $ in $
t\rightarrow\infty $ as
\begin{eqnarray}\label{d11.05}
&& b_{\vec{k}}(t)=b_{\vec{k}}(0)e^{-i\omega_{\vec{k}}
t}-i\sqrt{\frac{3\beta }{4\pi^2
\omega_{\vec{k}}^5}}\frac{e^{-i\omega_{\vec{k}}
t}}{(\omega^2-\omega_{\vec{k}}^2-\frac{i\beta}{m}\omega_{\vec{k}})}\{\omega^2
\vec{x}(0)+i\omega_{\vec{k}}\frac{\vec{p}(0)-\vec{R}(0)}{m}\nonumber\\
&&-\vec{M}(0)\omega^2-i\omega_{\vec{k}}\dot{\vec{M}}(0)\}\nonumber\\
&&+\frac{3\beta i}{4\pi^2m\sqrt{\omega_{\vec{k}}^5}}
\int_{-\infty}^{+\infty}\frac{d^3k'}{
\sqrt{\omega_{\vec{k'}}}}\{\frac{b_{\vec{k'}}(0)}{\omega^2-\omega_{\vec{k'}}^2-\frac{i\beta}{m}\omega_{\vec{k'}}}
\frac{\sin\frac{(\omega_{\vec{k}}-\omega_{\vec{k'}}t)
}{2}t}{\frac{(\omega_{\vec{k}}-\omega_{\vec{k'}})}{2}}e^{\frac{-i(\omega_{\vec{k}}+\omega_{\vec{k'}})t}{2}}\nonumber\\
&&+\frac{b_{\vec{k'}}^\dag(0)}{\omega^2-\omega_{\vec{k'}}^2+\frac{i\beta}{m}\omega_{\vec{k'}}}
\frac{\sin\frac{(\omega_{\vec{k}}+
\omega_{\vec{k'}})}{2}t}{\frac{(\omega_{\vec{k}}+\omega_{\vec{k'}})}{2}}
e^{\frac{i(\omega_{\vec{k'}}-\omega_{\vec{k}})t}{2}}\}\vec{k}.\vec{k'},
\end{eqnarray}
now substituting $ b_{\vec{k}}(t) $ from (\ref{d11.05}) in
(\ref{d5})
and using (\ref{d6}), one obtains $ \vec{p}=m\dot{\vec{x}}+\vec{R} $.\\
A vector in fock space of reservoir is a linear combination of
basis vectors
\begin{equation}\label{d15.121}
|N(\vec{k}_1),N(\vec{k}_2),...\rangle_B=\frac{(b_{\vec{k}_1})^{N(\vec{k}_1)}(b_{\vec{k}_2})^{N(\vec{k}_2)}...}{\sqrt{N(\vec{k}_1)!N(\vec{k}_2)!...}}|0\rangle_B
\end{equation}
 where are eigenstates of $ H_B $ and the operators $ b_{\vec{k}} $ and $ b_{\vec{k}}^\dag $ act on them as
\begin{eqnarray}\label{d15.131}
&&b_{\vec{k}}|N(\vec{k}_1),N(\vec{k}_2),...N(\vec{k}),...\rangle_B=\sqrt{N(\vec{k})}|N(\vec{k}_1),N(\vec{k}_2),...N(\vec{k})-1,...\rangle_B\nonumber\\
&&b_{\vec{k}}^\dag|N(\vec{k}_1),N(\vec{k}_2),...N(\vec{k}),...\rangle_B=\sqrt{N(\vec{k})+1}|N(\vec{k}_1),N(\vec{k}_2),...N(\vec{k})+1,...\rangle_B\nonumber\\
&&
\end{eqnarray}
If the state of system in $ t=0 $ is taken to be $
|\psi(0)\rangle=|0\rangle_B\otimes|n_1,n_2,n_3\rangle_\omega  $
where $ |0\rangle_B $ is the vacuum state of the reservoir and $
|n_1,n_2,n_3\rangle_\omega $ is an excited state of the
Hamiltonian $
H_S=\frac{\vec{p}^2}{2m}+\frac{1}{2}m\omega^2\vec{x}^2 $, then it
is clear that
\begin{equation}\label{d11.06}
\langle\psi(0)|H_s(0)|\psi(0)\rangle=(n_1+n_2+n_3+\frac{3}{2})\omega.
\end{equation}
 On the other hand from (\ref{d11.04}), (\ref{d11.05}) and (\ref{d5})
we find
\begin{eqnarray}\label{d11.07}
&& lim_{t\rightarrow\infty}[ \langle\psi(0)| :
 \frac{1}{2}m\dot{\vec{x}}^2+\frac{1}{2}m\omega^2 \vec{x}^2 :|\psi(0)\rangle]= 0,\nonumber\\
&&lim_{t\rightarrow\infty}[ \langle\psi(0)| : H_s(t) :|\psi(0)\rangle]\nonumber\\
&&=\frac{\beta^2\omega^4}{2\pi^2
m}lim_{t\rightarrow\infty}|\int_{-\infty}^{+\infty}\frac{d
x}{x}\frac{e^{ixt}}{\omega^2-x^2+i\frac{\beta}{m}x}|^2\langle n_1,n_2,n_3|\vec{x}^2(0)|n_1,n_2,n_3\rangle\nonumber\\
&&\simeq\frac{\beta^2}{2m}\langle n_1,n_2,n_3
\vec{x}^2(0)|n_1,n_2,n_3\rangle=\frac{\beta^2}{2m^2\omega}(n_1+n_2+n_3+\frac{3}{2}),
\end{eqnarray}
where  $:\hspace{01.00 cm}  : $ denotes the normal ordering
operator. Now by substituting $ b_{\vec{k}}(t) $ from
(\ref{d11.05}) into (\ref{d3}), it is easy to show that
\begin{eqnarray}\label{d11.08}
&&lim_{t\rightarrow\infty}[ \langle\psi(0)|: H_B(t) :|\psi(0)\rangle]\nonumber\\
&&=\frac{\beta\omega^4}{\pi}\int_0^\infty \frac{d
x}{(\omega^2-x^2)^2+\frac{\beta^2}{m^2}x^2}\langle n_1,n_2,n_3|
\vec{x}^2(0)|n_1,n_2,n_3\rangle+\nonumber\\
&&+\frac{\beta}{\pi m^2 }\int_0^\infty \frac{x^2d x
}{(\omega^2-x^2)^2+\frac{\beta^2}{m^2}x^2}\langle n_1,n_2,n_3|
\vec{p}^2(0)|n_1,n_2,n_3\rangle \nonumber\\
&&=\frac{\beta\omega^3}{\pi
m}(n_1+n_2+n_3+\frac{3}{2})\int_0^\infty \frac{d
x}{(\omega^2-x^2)^2+\frac{\beta^2}{m^2}x^2}+\nonumber\\
&&+\frac{\beta\omega}{\pi m}(n_1+n_2+n_3+\frac{3}{2})\int_0^\infty
\frac{x^2d
x}{(\omega^2-x^2)^2+\frac{\beta^2}{m^2}x^2}.\nonumber\\
&&
\end{eqnarray}
For sufficiently weak damping that is when $\beta $ is very small
, the integrands in (\ref{d11.08}) have singularity points $
x=\pm(\omega_1\pm\frac{i\beta}{2m})$ and by using  residual
calculus we find
\begin{equation}\label{d11.09}
lim_{t\rightarrow\infty}[ \langle\psi(0)|: H_B :|\psi(0)\rangle]
=(n_1+n_2+n_3+\frac{3}{2})\omega,
\end{equation}
Comparing (\ref{d11.06}) and (\ref{d11.09}), show that the total
energy of oscillator has been transmited to the reservoir and
according to (\ref{d11.07}), the kinetic energy of oscillator
tends to zero.\\
If the state of system  in $ t=0 $ is $
\rho(0)=\rho_B^T\otimes|S\rangle_\omega $ where $
\rho_B^T=\frac{e^{\frac{-H_B}{KT}}}{Tr_B(e^{\frac{-H_B}{KT}})} $
is the Maxwell-Boltzman distribution and $ |S\rangle_\omega $ is
an arbitrary state of harmonic oscillator , then by making use of
$
Tr_B[b_{\vec{k}}^\dag(0)b_{\vec{k'}}(0)\rho_B^T]=\frac{\delta(\vec{k}-\vec{k'})}{e^{\frac{\omega_{\vec{k}}}{KT}}-1}$
one can show that
\begin{eqnarray}\label{d11.1}
&&lim_{t\rightarrow\infty} \langle
:\frac{1}{2}m\dot{\vec{x}}^2(t)+\frac{1}{2}m\omega^2 \vec{x}^2(t)
: \rangle =\frac{6\beta}{\pi m^2}\int_0^\infty
\frac{x}{[(\omega^2-x^2)^2+\frac{\beta^2}{m^2}x^2](e^{\frac{x}{KT}}-1)}dx\nonumber\\
&&+\frac{6\beta}{\pi m^2}\int_0^\infty
\frac{x^3}{[(\omega^2-x^2)^2+\frac{\beta^2}{m^2}x^2](e^{\frac{x}{KT}}-1)}dx.
\end{eqnarray}
\subsection{Transition probabilities}
We can write the Hamiltonian (\ref{d1}) as
\begin{eqnarray}\label{d23}
&&H=H_0+H',\nonumber\\
&&H_0=\sum_{j=1}^3(a_j^\dag a_j+\frac{3}{2})\omega+H_B, \nonumber\\
&&H'=-\frac{\vec{p}}{m}.\vec{R}+\frac{\vec{R}^2}{2m},
\end{eqnarray}
 where $ a_j , a_j^\dag \hspace{.50 cm}j=1,2,3 $ are annihilation and creation operators of
 the harmonic oscillator.
 In interaction picture we can write
\begin{eqnarray}\label{d24}
&& a_{jI}(t)=e^{iH_0 t}a_j(0)e^{-iH_0 t}=a_j(0)e^{-i\omega t}\hspace{1.00 cm} j=1,2,3,\nonumber\\
&& b_{\vec{k}I}(t)=e^{iH_0 t}b_{\vec{k}}(0)e^{-iH_0
t}=b_{\vec{k}}(0)e^{-i\omega_{\vec{k}} t},
\end{eqnarray}
the terms $ \frac{\vec{R}}{m}.\vec{ p} $ and $
\frac{\vec{R}^2}{2m} $ are of the first order and second order of
damping respectively, therefore, for a sufficiently weak damping,
$ \frac{\vec{R}^2}{2m}$ is small in comparison with $
\frac{\vec{R}}{m}.\vec{p} $. Furthermore  $ \frac{\vec{R}^2}{2m}
$ has not any role in those transition probabilities where
initial and final states of harmonic oscillator are different,
hence we can neglect it in $ H' $. Substituting $ a_{jI}
\hspace{1.00 cm}j=1,2,3  $ and $ b_{\vec{k}I} $ from (\ref{d24})
in $\frac{\vec{R}}{m}.\vec{ p} $, one can obtain $ H_I' $ in
interaction picture as
\begin{eqnarray}\label{d25}
&&H_I'=-i\sqrt{\frac{\omega}{2m}}\sum_{j=1}^3\int_{-\infty}^{+\infty}d^3
k [f(\omega_{\vec{k}}) a_j^\dag b_{\vec{k}}(0)
e^{i(\omega-\omega_{\vec{k}})t}-f^*(\omega_{\vec{k}}) a_j
b_{\vec{k}}^\dag(0) e^{-i(\omega-\omega_{\vec{k}}
)t}\nonumber\\
&&-f(\omega_k) a_j b_{\vec{k}}(0)
e^{-i(\omega_{\vec{k}}+\omega)t}+f^*(\omega_k) a_j^\dag
b_{\vec{k}}(0) e^{i(\omega_{\vec{k}}+\omega) t}]k_j,
\end{eqnarray}
 where $ k_j \hspace{0.50 cm} j=1,2,3 $ are the cartesian  companent of the vector $ \vec{k} $.
 The terms  containing just $
a_j b_{\vec{k}}(0) $ and $ a_j^\dag b_{\vec{k}}^\dag(0) $ violate
the conservation of energy in the first order perturbation,
because $ a_j b_{\vec{k}}(0) $ destroys an excited state of
harmonic oscillator while at the same time destroying a reservoir
excitation state and $ a_j^\dag b_{\vec{k}}^\dag(0) $ creates an
excited state of harmonic oscillator, while creating an excited
reservoir state at the same time, therefore we neglect them due to
energy conservation and write $ H'_I $ as
\begin{equation}\label{d26}
H_I'=-i\sqrt{\frac{\omega}{2m}}\sum_{j=1}^3\int_{-\infty}^{+\infty}d^3
k [f(\omega_{\vec{k}}) a_j^\dag b_{\vec{k}}(0)
e^{i(\omega-\omega_{\vec{k}})t}-f^*(\omega_{\vec{k}}) a_j
b_{\vec{k}}^\dag(0) e^{-i(\omega-\omega_{\vec{k}} )t}]k_j.
\end{equation}
The time evolution of density operator in interaction picture is
[14]
\begin{equation}\label{d27}
\rho_I(t)=U_I(t,t_0)\rho_I(t_0)U_I^\dag(t,t_0),
\end{equation}
where $ U_I $ is the time evolution operator, which in first
order perturbation is
\begin{eqnarray}\label{d28}
 &&U_I(t,t_0=0)=1-i \int_0^t d t_1 H'_I(t_1)=\nonumber\\
 &&= 1-\sqrt{\frac{\omega}{2m}}\sum_{j=1}^3 \int_{-\infty}^{+\infty}d^3 k
k_j[f(\omega_{\vec{k}}) a_j^\dag b_{\vec{k}}(0)
e^{\frac{i(\omega-\omega_{\vec{k}})
t}{2}}\nonumber\\
&&-f^*(\omega_{\vec{k}}) a_j b_{\vec{k}}^\dag(0)
e^{\frac{-i(\omega-\omega_{\vec{k}})t}{2}}]
\frac{\sin\frac{(\omega-\omega_{\vec{k}})}{2}t}{\frac{(\omega-\omega_{\vec{k}})}{2}}.
\end{eqnarray}
Let $ \rho_I(0)=|n_1,n_2,n_3\rangle_{\omega\hspace{0.20
cm}\omega}\langle n_1,n_2,n_3|\otimes|0\rangle_{B\hspace{0.20
cm}B} \langle
 0|$ where $ |0\rangle_B $ is the vacuum state of the reservoir and
 $ |n_1,n_2,n_3\rangle_\omega $, is an
 excited state of the harmonic oscillator, substituting $ U_I(t,0) $ from
 (\ref{d28}) in (\ref{d27}) and tracing out the reservoir parameters, we
 obtain the probability that the reservoir absorbs energy $ \omega $ from
 oscillator
 \begin{eqnarray}\label{d28.1}
&&\Gamma_{(n_1+n_2+n_3+\frac{3}{2})\omega\rightarrow(
n_1+n_2+n_3+\frac{1}{2})\omega}=
Tr_s[(|n_1-1,n_2,n_3\rangle_{\omega\hspace{0.20cm}\omega}\langle
|n_1-1,n_2,n_3|+\nonumber\\
&&+|n_1,n_2-1,n_3\rangle_{\omega\hspace{0.20 cm}\omega}\langle
|n_1,n_2-1,n_3|+ |n_1,n_2,n_3-1\rangle_{\omega\hspace{0.20
cm}\omega}\langle
|n_1,n_2,n_3-1|)\rho_{sI}(t)]=\nonumber\\
&&=\frac{2\pi\omega(n_1+n_2+n_3)}{3m}\int_{0}^{+\infty}
\omega_{\vec{k}}^4 |f(\omega_{\vec{k}})|^2
\frac{\sin^2\frac{(\omega_{\vec{k}}-\omega)}{2}t}
{(\frac{\omega_{\vec{k}}-\omega}{2})^2},
\end{eqnarray}
where  $ Tr_s $ denotes taking trace over harmonic oscillator and
$ \rho_{sI}(t) $ is obtained by taking trace of $ \rho_I(t) $ over
reservoir parameters i,e $ \rho_{sI}(t) =Tr_B(\rho_I(t)) $ and we
have used the formula $ Tr_B [ |1_{\vec{k}}\rangle_B
\hspace{00.20cm}_B\langle1_{\vec{k'}}| ]=\delta(\vec{k}-\vec{k'})$.\\
For very large times, we can write
$\frac{\sin^2\frac{(\omega_{\vec{p}}-\omega)}{2}t}{(\frac{\omega_{\vec{p}}-\omega}{2})^2}=2\pi
t\delta(\omega_{\vec{p}}-\omega)$ which leads to
\begin{equation}\label{d28.2}
\Gamma_{(n_1+n_2+n_3+\frac{3}{2})\omega\rightarrow(
n_1+n_2+n_3+\frac{1}{2})\omega}=\frac{4\pi^2 \omega^5(
n_1+n_2+n_3) t |f(\omega)|^2}{3m}.
\end{equation}
For the special choice (\ref{d10}), above transition probability
becomes
\begin{equation}\label{d30}
\Gamma_{(n_1+n_2+n_3+\frac{3}{2})\omega\rightarrow(
n_1+n_2+n_3+\frac{1}{2})\omega}=\frac{(n_1+n_2+n_3)\beta t}{m},
\end{equation}
 in this case the oscillator can ont absorb energy from reservoir. \\
 Now consider the case where the reservoir is an excited state in $ t=0$ for example $
\rho_I(0)=|n_1,n_2,n_3\rangle_{\omega\hspace{0.20
cm}\omega}\langle
n_1,n_2,n_3|\otimes|1_{\vec{p}_1},...1_{\vec{p}_r}\rangle_{B\hspace{0.20
cm}B} \langle
 1_{\vec{p}_1},...1_{\vec{p}_r}| $ where $ |1_{\vec{p}_1},...1_{\vec{p}_r}\rangle_B $ denotes a state of
 reservoir that contains $ r $ quanta with corresponding momenta $ \vec{p}_1,...\vec{p}_r $,
 then
 \begin{eqnarray}\label{d31}
 &&Tr_B[ b_{\vec{k}}^\dag |1_{\vec{p}_1},...1_{\vec{p}_r}\rangle_{B\hspace{0.20 cm}B} \langle
 1_{\vec{p}_1},...1_{\vec{p}_r}| b_{\vec{k'}}]=\delta(\vec{k}-\vec{k'}),\nonumber\\
 &&Tr_B[b_{\vec{k}} |1_{\vec{p}_1},...1_{\vec{p}_r}\rangle_{B\hspace{0.20 cm}B} \langle
 1_{\vec{p}_1},...1_{\vec{p}_r}|b_{\vec{k'}}^\dag ]=
 \sum_{l=1}^r \delta(\vec{k}-\vec{p}_l)\delta(\vec{k'}-\vec{p}_l),
 \end{eqnarray}
 and from the long time approximation, we find
\begin{eqnarray}\label{d32}
&&\Gamma_{(n_1+n_2+n_3+\frac{3}{2})\omega\rightarrow(
n_1+n_2+n_3+\frac{1}{2})\omega}=\frac{4\pi^2
\omega^5( n_1+n_2+n_3) t}{3m}|f(\omega)|^2,\nonumber\\
&&\Gamma_{(n_1+n_2+n_3+\frac{3}{2})\omega\rightarrow(
n_1+n_2+n_3+\frac{5}{2})\omega}=\nonumber\\
&&\frac{\omega \pi t}{m}|f(\omega)|^2 \sum_{l=1}^r \delta(
\omega_{\vec{p}_l}-\omega)[(n_1+1)p_{l1}^2+(n_2+1)p_{l2}^2+(n_3+1)p_{l3}^2],\nonumber\\
&&
\end{eqnarray}
where $ p_{l1},p_{l2},p_{l3} $ are the cartesian components of
vector $ \vec{p}_l $ . Specially for the choice (\ref{d10}), we
have
\begin{eqnarray}\label{d33}
&&\Gamma_{(n_1+n_2+n_3+\frac{3}{2})\omega\rightarrow(
n_1+n_2+n_3+\frac{1}{2})\omega}=\frac{(n_1+n_2+n_3)\beta t}{m},\nonumber\\
&&\Gamma_{(n_1+n_2+n_3+\frac{3}{2})\omega\omega\rightarrow(
n_1+n_2+n_3+\frac{5}{2})\omega}=\nonumber\\
&&=\frac{3\beta t}{4\pi m \omega^4} \sum_{l=1}^r \delta(
\omega_{\vec{p}_l}-\omega)[(n_1+1)p_{l1}^2+(n_2+1)p_{l2}^2+(n_3+1)p_{l3}^2]\nonumber\\
&&.
\end{eqnarray}
Another important case is when the reservoir has a
Maxwell-Boltzman distribution, so let $\rho_I(0)=
|n_1,n_2,n_3\rangle_{\omega\hspace{0.20 cm}\omega}\langle
n_1,n_2,n_3|\otimes \rho_B^T $ where\\  $
\rho_B^T=\frac{e^{\frac{-H_B}{K T}}}{TR_B(e^{\frac{-H_B}{KT}})}
$, then using the relations
\begin{eqnarray}\label{d34}
&&Tr_B[ b_{\vec{k}}\rho_B^T b_{\vec{k'}}]=Tr_B[ b_{\vec{k}}^\dag
\rho_B^T
b_{\vec{k'}}^\dag]=0,\nonumber\\
&& Tr_b[b_{\vec{k}}\rho_B^T
b_{\vec{k'}}^\dag]=\frac{\delta(\vec{k}-\vec{k'})}
{e^{\frac{\omega_{\vec{k}}}{K T}}-1},\nonumber\\
&&Tr_B[ b_{\vec{k}}^\dag \rho_B^T
b_{\vec{k'}}]=\frac{\delta(\vec{k}-\vec{k'})e^{\frac{\omega_{\vec{k}}}{K
T}}}{e^{\frac{\omega_{\vec{k}}}{K T}}-1},
\end{eqnarray}
we can obtain the following transition probabilities in very long
time
\begin{eqnarray}\label{d35}
&& \Gamma_{(n_1+n_2+n_3+\frac{3}{2})\omega\rightarrow(
n_1+n_2+n_3+\frac{1}{2})\omega}=\frac{4\pi^2 \omega^5(
n_1+n_2+n_3) t|}{3m}\frac{f(\omega)|^2e^{\frac{\omega}{K
T}}}{e^{\frac{\omega}{K T}}-1},\nonumber\\
&&\Gamma_{(n_1+n_2+n_3+\frac{3}{2})\omega\rightarrow(
n_1+n_2+n_3+\frac{5}{2})\omega}=\nonumber\\
 &&=\frac{4\pi^2\omega^5
(n_1+n_2+n_3+3)
t}{3m}\frac{|f(\omega)|^2}{e^{\frac{\omega}{K T}}-1},\nonumber\\
&&
\end{eqnarray}
substituting (\ref{d10}) in these recent relations we find
\begin{eqnarray}\label{d36}
&&\Gamma_{(n_1+n_2+n_3+\frac{3}{2})\omega\rightarrow(
n_1+n_2+n_3+\frac{1}{2})\omega}=\frac{( n_1+n_2+n_3)\beta
te^{\frac{\omega}{K
T}}}{m(e^{\frac{\omega}{K T}}-1)},\nonumber\\
 &&\Gamma_{(n_1+n_2+n_3+\frac{3}{2})\rightarrow(
n_1+n_2+n_3+\frac{5}{2})}=\frac{(n_1+n_2+n_3+3)\beta
t}{m(e^{\frac{\omega}{K T}}-1)}.\nonumber\\
&&
\end{eqnarray}
 So in very low temperatures the energy flows from oscillator to the reservoir by the rate
  $ \Gamma_{(n_1+n_2+n_3+\frac{3}{2})\omega\rightarrow(
n_1+n_2+n_3+\frac{1}{2})\omega}=\frac{( n_1+n_2+n_3)\beta }{m},$
and no energy flows from the reservoir to the oscillator.
\section{A dissipative two level quantum system}
Let us write the Hamiltonian (\ref{d1}) as
\begin{equation}\label{d12}
H=H_s+H_B-\frac{\vec{R}}{m}.\vec{p}+\frac{\vec{R}^2}{2m},
\end{equation}
where $ H_S=\frac{\vec{p}^2}{2m}+v(\vec{x}) $ is the Hamiltonian
of the system. The $ H_s $ has two eigenvalues $ E_1, E_2 $
corresponding to eigenkets $ |1\rangle, |2\rangle $ respectively,
and we can write
\begin{equation}\label{d13}
H_s=\frac{1}{2}(E_2-E_1)\sigma_z+\frac{1}{2}(E_1+E_2)\hspace{1.50
cm}\sigma_z=|2\rangle \langle2|-|1\rangle\langle1|.
\end{equation}
  For sufficiently weak damping we can neglect the term $
\frac{\vec{R}^2}{2m}$, because it is proportional to the second
order of damping and write $ H $ as
\begin{eqnarray}\label{d14}
&&H=\frac{1}{2}\omega_0\sigma_z+H_B-\int d^3k [( G_{12}\sigma
+G_{21}\sigma^\dag ]b_{\vec{k}}-\int d^3k b_{\vec{k}}^\dag[(
G_{21}^*\sigma +G_{12}^*\sigma^\dag ]\nonumber\\
&& G_{ij}=\frac{f( \omega_{\vec{k}})}{m}
\vec{k}.\vec{p}_{ij}=i\omega_{ij}f(
\omega_{\vec{k}})\vec{k}.\vec{x}_{ij}\nonumber\\
&&\vec{x}_{ij}=\langle i|\vec{x}|j\rangle \hspace{1.50
cm}\vec{p}_{ij}=\langle i|\vec{p}|j\rangle \hspace{1.00
cm}i,j=1,2,
\end{eqnarray}
where $ \omega_0=E_2-E_1 $  and $ \sigma=|1\rangle\langle 2| $.
Using commutation relations
\begin{equation}\label{d15}
[\sigma,\sigma^\dag]=-\sigma_z \hspace{1.00 cm}
[\sigma,\sigma_z]=2\sigma \hspace{1.00
cm}[\sigma_z,\sigma^\dag]=2\sigma^\dag,
\end{equation}
one can easily obtain Heisenberg equations for the two level
system
\begin{eqnarray}\label{d16}
&&\dot{\sigma}=i[H,\sigma]=-i\omega_0\sigma-i\int d^3k[
G_{21}\sigma_z
b_{\vec{k}}+G_{12}^*b_{\vec{k}}^\dag\sigma_z]\nonumber\\
&&\dot{\sigma}_z=i[H,\sigma_z]=-2i\int d^3k[ G_{12}\sigma
b_{\vec{k}}-G_{21}\sigma^\dag b_{\vec{k}}]-2i\int d^3k[ G_{21}^*
b_{\vec{k}}^\dag \sigma-G_{12}^* b_{\vec{k}}^\dag\sigma^\dag].\nonumber\\
&&
\end{eqnarray}
Since equal-time system and reservoir operators commute, we can
write the Heisenberg equations (\ref{d16}) in different but
equivalent ways. For example we can use the normal ordering were
annihilation operator of the reservoir $ b_{\vec{k}} $ appears at
the right and creation operator $ b_{\vec{k}}^\dag $ appears at
the left of the system operators, i.e, $
\sigma,\sigma^\dag,\sigma_z $ , or we can use antinormal ordering
wereh annihilation operator of the reservoir $ b_{\vec{k}} $
appears at the left and creation operator $ b_{\vec{k}}^\dag $
appears at the right of the system operators, i.e, $
\sigma,\sigma^\dag,\sigma_z $. In equations (\ref{d16}) we have
used the normal ordering.\\
 One can easily obtain the Heisenberg equation for $ b_{\vec{k}} $ as
\begin{equation}\label{d17}
\dot{b}_{\vec{k}}=i[H,b_{\vec{k}}]=-i\omega_{\vec{k}}
b_{\vec{k}}+i\int d^3k G_{21}^*\sigma+i\int d^3k
G_{12}^*\sigma^\dag,
\end{equation}
with the following formal solution
\begin{equation}\label{d18}
b_{\vec{k}}(t)=b_{\vec{k}}(0)e^{-i\omega_{\vec{k}}
t}+iG_{21}^*\int_0^t dt'
\sigma(t')e^{-i\omega_{\vec{k}}(t-t')}+iG_{12}^*\int_0^t dt'
\sigma^\dag(t')e^{-i\omega_{\vec{k}}(t-t')}.
\end{equation}
Let us assume that damping is  sufficiently weak that we can take
the itegrands (\ref{d18}) we can take
\begin{equation}\label{d19}
\sigma(t')\cong\sigma(t)e^{-i\omega_0(t'-t)}\hspace{1.50
cm}\sigma^\dag(t')\cong\sigma^\dag(t)e^{i\omega_0(t'-t)}.
\end{equation}
This is called the Markovian approximation [15] which replaces the
system operators in (\ref{d18}) by an operator that depends on
the system operators at the same time $ t $, without taking into
account the memory of these operators at earlier times.
Substituting (\ref{d19}) in (\ref{d18}) and using (\ref{d16}) we
find
\begin{eqnarray}\label{d20}
&&\langle\dot{\sigma}\rangle=-i\omega_0\langle\sigma\rangle+\int
d^3k|G_{21}|^2[\langle\sigma_z\sigma\rangle\int_0^t dt'
e^{i(\omega_{\vec{k}}-\omega_0)(t'-t)}
-\langle\sigma\sigma_z\rangle \int_0^t dt'
e^{i(\omega_{\vec{k}}+\omega_0)(t'-t)}]\nonumber\\
&&+\int d^3k G_{21}
G_{12}^*[\langle\sigma_z\sigma^\dag\rangle\int_0^t dt'
e^{i(\omega_{\vec{k}}+\omega_0)(t'-t)}
-\langle\sigma^\dag\sigma_z\rangle \int_0^t dt'
e^{i(\omega_{\vec{k}}-\omega_0)(t'-t)}]\nonumber\\
&&\langle\dot{\sigma}_z\rangle=2\langle\sigma\sigma^\dag\rangle\int
d^3k|G_{12}|^2[\int_0^t dt' e^{i(\omega_{\vec{k}}+\omega_0)(t'-t)}
+\int_0^t dt'
e^{-i(\omega_{\vec{k}}+\omega_0)(t'-t)}]\nonumber\\
&&-2\langle\sigma^\dag\sigma \rangle\int d^3k |G_{21}|^2 [\int_0^t
dt' e^{i(\omega_{\vec{k}}-\omega_0)(t'-t)} + \int_0^t
dt'e^{-i(\omega_{\vec{k}}-\omega_0)(t'-t)}],\nonumber\\
&&
\end{eqnarray}
where we have taken expectation values in state $ |\psi(0)\rangle=
|0\rangle_B\otimes|S\rangle $ which is a tensor product of the
vacuum state of the reservoir $ |0\rangle_B $ and an arbitrray
state of the system, $ |S\rangle $. In long time approximation, we
can write
\begin{eqnarray}\label{d21}
&&\int d^3k|G_{21}|^2\int_0^t dt'
e^{i(\omega_{\vec{k}}-\omega_0)(t'-t)}=\nonumber\\
&&=\frac{4\pi\omega_0^2|\vec{x}_{21}|^2}{3}\int_0^\infty
d\omega_{\vec{k}}|f(\omega_{\vec{k}})|^2\omega_{\vec{k}}^4
[\frac{-i}{\omega_{\vec{k}}-\omega_0}+\pi\delta(\omega_{\vec{k}}-\omega_0)]\nonumber\\
&&\int d^3k|G_{21}|^2\int_0^t dt'
e^{i(\omega_{\vec{k}}+\omega_0)(t'-t)}=-i\frac{4\pi\omega_0^2|\vec{x}_{21}|^2}{3}\int_0^\infty
d\omega_{\vec{k}}
\frac{|f(\omega_{\vec{k}})|^2\omega_{\vec{k}}^4}{\omega_{\vec{k}}+\omega_0}\nonumber\\
&&\int d^3k G_{21} G_{12}^*\int_0^t dt'
e^{i(\omega_{\vec{k}}-\omega_0)(t'-t)}=\nonumber\\
&&=-\frac{4\pi\omega_0^2|\vec{x}_{21}|^2}{3}\int_0^\infty
d\omega_{\vec{k}}|f(\omega_{\vec{k}})|^2\omega_{\vec{k}}^4[\frac{-i}{\omega_{\vec{k}}-\omega_0}+\pi\delta(\omega_{\vec{k}}-\omega_0)]\nonumber\\
&&\int d^3k G_{21} G_{12}^*\int_0^t dt'
e^{i(\omega_{\vec{k}}+\omega_0)(t'-t)}=i\frac{4\pi\omega_0^2|\vec{x}_{21}|^2}{3}\int_0^\infty
d\omega_{\vec{k}}\frac{|f(\omega_{\vec{k}})|^2\omega_{\vec{k}}^4}
{\omega_{\vec{k}}+\omega_0},\nonumber\\
&&
\end{eqnarray}
substitution of ( \ref{d21}) in (\ref{d20}) gives
\begin{eqnarray}\label{d22}
&&\langle\dot{\sigma}_z\rangle=-4\mu\langle\sigma^\dag\sigma\rangle=-2\mu(1+\langle\sigma_z\rangle)\hspace{1.50
cm}\mu=\frac{4\pi^2\omega_0^6|f(\omega_0)|^2}{3}\nonumber\\
&&\langle\dot{\sigma}\rangle=-i\omega_0\langle\sigma\rangle+(i\triangle_1-i\triangle_2-\mu)\langle\sigma\rangle
+(i\triangle_2-i\triangle_1-\mu)\langle\sigma^\dag\rangle \nonumber\\
&&\triangle_1=\frac{4\pi^2\omega_0^6|\vec{x}_{12}|^2}{3}\int_0^\infty
d\omega_{\vec{k}}
\frac{|f(\omega_{\vec{k}})|^2\omega_{\vec{k}}^4}{\omega_{\vec{k}}-\omega_0}\hspace{1.50
cm}\triangle_2=\frac{4\pi^2\omega_0^6|\vec{x}_{12}|^2}{3}\int_0^\infty
d\omega_{\vec{k}}
\frac{|f(\omega_{\vec{k}})|^2\omega_{\vec{k}}^4}{\omega_{\vec{k}}+\omega_0}.\nonumber\\
&&
\end{eqnarray}
The solution of the first equation in (\ref{d22}) is
\begin{equation}\label{d23}
\langle\sigma_z(t)\rangle=-1+(1+\langle\sigma_z(0)\rangle)e^{-2\mu.
t}
\end{equation}
If $ \langle\sigma_z(0)\rangle=1 $ , i.e , the two level system is
initially in the upper state $ |2\rangle $, it decays to the
lower state $ \langle\sigma_z\rangle=-1 $  with the rate $ 2\mu
$, and if the system is initially in the lower state, it
remains for all times in that state. \\
By defining $ \hat{F}=\sigma+\sigma^\dag $ and $
\hat{E}=\sigma^\dag-\sigma $, the second equation in (\ref{d22})
can be written as
\begin{eqnarray}\label{d24.125}
&&\langle\dot{\hat{F}}\rangle=i\Gamma\langle\hat{E}\rangle-2\mu\langle\hat{F}\rangle\hspace{1.50
cm}\Gamma=\omega_0-2\triangle_2-2\triangle_1,\nonumber\\
&&\langle\dot{\hat{E}}\rangle=i\omega_0\langle\hat{F}\rangle,
\end{eqnarray}
with the following solutions
\begin{eqnarray}\label{d25}
&&\langle\hat{F}\rangle=\hat{C}_1e^{i\Omega_+
t}+\hat{C}_2e^{i\Omega_- t},\hspace{1.50cm}\Omega_{\pm}=i\mu \pm i\sqrt{\mu^2+\omega_0\Gamma}\nonumber\\
&&
\langle\hat{E}\rangle=\frac{\omega_0}{\Omega_+}\hat{C}_1e^{i\Omega_+
t}+\frac{\omega_0}{\Omega_-}\hat{C}_2e^{i\Omega_- t}.
\end{eqnarray}
If $ \mu^2+\omega_0\Gamma<0 $, we conclude that $
\langle\sigma\rangle $ decays with the rate $ \mu $.
\section{Quantum field of the reservoir}
Let us define the operators $ Y(\vec{x},t) $ and $
\Pi_Y(\vec{x},t) $ as follows
\begin{eqnarray}\label{d25.01}
&&Y(\vec{x},t)=\int_{-\infty}^{+\infty} \frac{d^3
k}{\sqrt{2(2\pi)^3\omega_{\vec{k}}}}(
b_{\vec{k}}(t)e^{i\vec{k}.\vec{x}}+b_{\vec{k}}^\dag(t)e^{-i\vec{k}.\vec{x}}),\nonumber\\
&&\Pi_Y(\vec{x},t)=i\int_{-\infty}^{+\infty} d^3 k
\sqrt{\frac{\omega_{\vec{k}}}{2(2\pi)^3}}( b_{\vec{k}}^\dag
(t)e^{-i\vec{k}.\vec{x}}-b_{\vec{k}}(t)e^{i\vec{k}.\vec{x}}),
\end{eqnarray}
then using commutation relations (\ref{d4}), one can show that $
Y(\vec{x},t) $ and $ \Pi_Y (\vec{x},t)$, satisfy the equal time
commutation relations
\begin{equation}\label{d25.02}
[ Y(\vec{x},t),\Pi_Y(\vec{x'},t)]=i\delta(\vec{x}-\vec{x'}),
\end{equation}
furthermore by substituting $ b_{\vec{k}}(t)$ from (\ref{d8}) in
(\ref{d25.01}), we obtain
\begin{eqnarray}\label{d25.03}
&&\frac{\partial\Pi_Y(\vec{x},t)}{\partial
t}=\nabla^2Y+2\dot{\vec{x}}(t). \vec{M}(\vec{x}),\hspace{1.00 cm
}\vec{M}(\vec{x})=Re{ \int_{-\infty}^{+\infty} d^3
k\sqrt{\frac{\omega_{\vec{k}}}{2(2\pi)^3}}f(\omega_{\vec{k}})\vec{k}
e^{-i\vec{k}.\vec{x}}},\nonumber\\
&&\Pi_Y(\vec{x},t)=\frac{\partial Y}{\partial
t}-2\dot{\vec{x}}(t) .\vec{N}(\vec{x}),\hspace{1.00 cm}
\vec{N}(\vec{x})= I m{\int_{-\infty}^{+\infty} d^3
k\frac{f(\omega_{\vec{k}})}{\sqrt{2(2\pi)^3\omega_{\vec{k}}}}\vec{k}e^{-i\vec{k}.\vec{x}}},\nonumber\\
&&
\end{eqnarray}
so $ Y(\vec{x},t)$ satisfies the following source included
Klein-Gordon equation
\begin{equation}\label{d25.04}
\frac{\partial^2Y}{\partial
t^2}-\nabla^2Y=2\ddot{\vec{x}}(t).\vec{N}(\vec{x})+2\dot{\vec{x}}(t).\vec{M}(\vec{x}),
\end{equation}
with the corresponding Lagrangian density
\begin{equation}\label{d25.05}
\pounds=\frac{1}{2}(\frac{\partial Y}{\partial
t})^2-\frac{1}{2}\vec{\nabla Y}.\vec{\nabla
Y}-2\dot{\vec{x}}.\vec{N}(\vec{x})\frac{\partial Y}{\partial
t}+2\dot{\vec{x}}.\vec{M}(\vec{x})Y.
\end{equation}
Therefore the reservoir is a massless Klein-Gordon field with
source $
2\ddot{\vec{x}}.\vec{N}(\vec{x})+2\dot{\vec{x}}.\vec{M}(\vec{x})$.
The Hamiltonian density for (\ref{d25.04}) is as follows
\begin{equation}\label{d25.06}
\aleph=\frac{(\Pi_Y+2\dot{\vec{x}}.\vec{N})^2}{2}+\frac{1}{2}|\vec{\nabla
Y}|^2-2\dot{\vec{x}}.\vec{M}Y ,
\end{equation}
and equations (\ref{d25.03}) are Heisenberg equations for $ Y $
and $ \Pi_Y$. If we obtain $ b_{\vec{k}} $ and $ b_{\vec{k}}^\dag
$ from (\ref{d25.01}) in terms of $ Y $ and $ \Pi_Y $  and
substitute them in $ H_B $ defined in (\ref{d3}), we find
\begin{equation}\label{d22.75}
H_B=\int_{-\infty}^{+\infty}d^3k \omega_{\vec{k}}b_{\vec{k}}^\dag
b_{\vec{k}}=\frac{\Pi_Y^2}{2}+\frac{1}{2}|\vec{\nabla Y }|^2.
\end{equation}
\section{Concluding remarks}
 By generalizing Caldeira-Legget model to a reservoir with
 continuous degrees of freedom, for example a Klein-Gordon field,
 a new minimal coupling method introduced which can be extended and applied to a
 large class of dissipative quantum systems consistently.
This method applied to an isotropic three dimensional quantum
damped harmonic oscillator and a dissipative two level system as
prototypes of important dissipative models. Some
 transition probabilities explaining the way energy flows between
 subsystems obtained. Choosing different coupling functions in
 (\ref{d5}) we could investigate another classes of dissipative systems.
\\
\\
\\

\end{document}